\documentclass[pra,showpacs,preprint,amsmath,amssymb]{revtex4}
\def\qed{\leavevmode\unskip\penalty9999 \hbox{}\nobreak\hfill
     \quad\hbox{\leavevmode  \hbox to.77778em{%
               \hfil\vrule   \vbox to.675em%
               {\hrule width.6em\vfil\hrule}\vrule\hfil}}
     \par\vskip3pt}
\usepackage{epsfig}
\usepackage{amsmath}

\begin{document}

\begin{center}\bf  von Neumann Measurement-related Matrices and the Nullity Condition for Quantum Correlation \end{center}
\vskip 1mm

\begin{center}
Ming-Jing Zhao$^{1}$, {Teng Ma$^{2}$}, {Tinggui Zhang$^{3}$, {Shao-Ming Fei$^{4,5}$}
} \vspace{2ex}

\begin{minipage}{5in}

{\small $~^{1}$ School of Science,
Beijing Information Science and Technology University, Beijing, 100192, China}

{\small $~^{2}$ State Key Laboratory of Low-Dimensional Quantum Physics and Department of Physics, Tsinghua University, Beijing 100084, China}

{\small $~^{3}$ School of Mathematics and Statistics, Hainan
Normal University, Haikou, 571158, China}

\small $~^{4}$ {School of Mathematical Sciences, Capital Normal
University, Beijing 100048, China}

{\small $~^{5}$ Max-Planck-Institute for Mathematics in the
Sciences, 04103, Leipzig, Germany}

\end{minipage}

\vskip 4mm
\end{center}

\vskip 2mm
\parbox{14cm}
{{\bf Abstract} We study von Neumann measurement-related matrices, and the nullity condition of quantum correlation. We investigate the properties of these matrices that are related to a von Neumann measurement. It is shown that these
$(m^2-1)\times (m^2-1)$ matrices are idempotent, and have rank $m-1$. These properties give rise to necessary conditions for the nullity of quantum correlations in bipartite systems. Finally, as an example we discuss quantum correlation in Bell diagonal states.}



\section{Introduction}

A measurement is an ``irreversible" and ``non-deterministic" process, because once a measurement is completed and the information is extracted, the initial state is collapsed. Prior to a measurement, one cannot predict  with certainty either the measurement result or the post-measurement state. A von Neumann measurement is not the only type of measurement that one can perform on a quantum state. However, one of the reasons for its importance is that one can derive all other measurements from a von Neumann measurement.

In Ref. \cite{L. Roa}, it is shown that a known initial quantum state can be driven onto a known pure state without resorting to unitary transformations, by means of a discrete sequence consisting only of von Neumann measurements. Performing a von Neumann measurement on a subsystem of a composite quantum system can also result in unavoidable distillable entanglement \cite{M. A. Nielsen,Cao,Wulayimu} between the measurement apparatus and the system if the state has a nonzero quantum discord \cite{A. Streltsov}. Furthermore, von Neumann measurements can be used to classify quantum states in a bipartite system \cite{S. Luo2008}. A quantum state is classical-quantum if it is invariant under some von Neumann measurement ${\cal M}$ that acts on the first subsystem, i.e., $({\cal M}\otimes I)\rho ({\cal M}^\dagger\otimes I)=\rho$. Similarly, a state is called quantum-classical if $(I \otimes {\cal M}^\prime) \rho (I \otimes {\cal M}^{\prime\dagger})=\rho$ for some von Neumann measurement ${\cal M}^\prime$. Thus, quantum-classical and classical-quantum states are classical in one part. If a quantum state is classical in both two parts, then it is called
classical-classical. That is, such a state is invariant under some von Neumann measurements ${\cal M}$ and ${\cal M}^\prime$ acting on the first and second subsystems, respectively, $({\cal M} \otimes {\cal M}^\prime) \rho ({\cal M} \otimes {\cal M}^{\prime\dagger})=\rho$.
If a quantum state is not classical on either side, then it is quantum-quantum correlated. It is known that quantum-quantum states play a significant role in many information processing tasks and physical
processes.

Quantum correlation can be measured in various manners, such as quantum discord \cite{H. Ollivier, L. Henderson}, geometric quantum discord \cite{B. Dakic,S. Luo}, quantum deficit \cite{J. Oppenheim}, measurement induced disturbance \cite{S. Luo2008}, and relative entropy of discord \cite{K. Modi}. For example, quantum discord is the minimal amount of information remaining after measuring one of the subsystems. The quantum deficit is the difference between work that is extractable from the total system and the subsystems after suitable local operations and classical communications. Measurement induced disturbance refers to the loss of quantum mutual information following local projective measurements. Relative entropy of discord and geometric quantum discord are measures of correlation based on a distance metric. All of these measures have their own advantages for describing quantum correlation.

In practical terms, it is often only necessary to ascertain the nullity of quantum correlation or classicality, and the precise values of various measures are less important \cite{modi}. Classicality corresponds to the condition that maximal information regarding a subsystem can be obtained by some specific local measurement, without altering correlations with the rest of the system. Thus, it has applications in the theory of decoherence, where it describes the classical correlation between the pointer states of some measurement apparatus and the internal quantum states. The set of classical states can be used to define quantum correlation measures, using a notion of minimum distance as the relative entropy of discord and geometric quantum discord. Therefore, much effort has been contributed to describing the nullity of quantum correlation \cite{A. Ferraro,B. Dakic,A. Datta,L. Chen}. In Ref. \cite{B. Dakic}, a nullity condition was provided that makes use of the singular value decomposition.
In Ref. \cite{A. Ferraro}, a necessary condition for the nullity of a quantum correlation is presented in terms of the commutativity of quantum  states and reduced states.
Experimentally, practical entanglement witnesses are constructed based on a single quantum state \cite{J. Maziero} or multiple copies of a quantum state \cite{S. Yu, C. Zhang}, in order to detect the quantum correlation.

In this paper, we study the properties of matrices related to von Neumann measurements and the nullity conditions for quantum correlation.
Suppose that a von Neumann measurement acts on an $m$-dimensional quantum system. Then, its action on the basis composed of $m^2-1$ orthonormal traceless Hermitian
matrices corresponds to an $(m^2-1)\times (m^2-1)$ matrix. We show that this matrix is of rank $m-1$.
Based on this property, necessary conditions are derived for the nullity of quantum correlations in bipartite systems. Finally, some examples are analyzed to identify classicality in quantum states.

\section{von Neumann measurement-related matrices}

Let ${\cal M}$ be an arbitrary von Neumann measurement acting on an $m$-dimensional quantum system, ${\cal M}=\{|\phi_i\rangle \langle\phi_i|\}_{i=0}^{m-1}$.
Here, $|\phi_i\rangle=\sum_{j=0}^{m-1} a_{ij}|j\rangle$, and owing to the orthogonality and completeness of ${\cal M}$, $A=(a_{ij})$ is an $m\times m$ unitary matrix.
Under a von Neumann measurement, a quantum state $\rho$ is mapped to
${\cal M}(\rho)\equiv {\cal M}\rho {\cal M}^\dagger=\sum_{i=0}^{m-1} |\phi_i\rangle \langle \phi_i|\rho |\phi_i\rangle \langle \phi_i|$.
Suppose that $\{\mu_i\}_{i=1}^{m^2-1}$ is an orthonormal basis of the $m\times m$ traceless Hermitian matrices. Then, there exists an $(m^2-1)\times (m^2-1)$ matrix $M$ corresponding to ${\cal M}$ such that
\begin{equation}\label{operation and matrix}
\begin{array}{rcl}
&&{\cal M}(\mu_1,\mu_2,\ldots,\mu_{m^2-1})\\[1mm]
&\equiv&({\cal M}\mu_1{\cal M}^\dagger, {\cal M}\mu_2{\cal M}^\dagger, \ldots, {\cal M}\mu_{m^2-1}{\cal M}^\dagger)\\[1mm]
&=&(\mu_1,\mu_2,\ldots,\mu_{m^2-1})M,
\end{array}
\end{equation}
where $M$ is defined by ${\cal M}\mu_i{\cal M}^\dagger=\sum_{j=1}^{m^2-1} \mu_j M_{ji}$, with $i=1,\ldots, m^2-1$.
It is easy to prove that the matrix $M$ satisfies the condition that
$M^2=M$. Hence, $M$ is idempotent. In the following, we study the properties of the matrix $M$, and show that $M$ is also singular.
Before we present our main results, we require the following lemma.

{\bf Lemma}~Let $A$ be an arbitrary $m\times m$ unitary matrix, with the $i$-th row and $j$-th column entry given by $a_{i-1,j-1}$.
Let $C_1$ be an $m\times (m-1)$ matrix with the $s$-th row and $p$-column entry given by
$\sqrt{\frac{1}{p(p+1)}}[\sum_{j=0}^{p-1} |a_{s-1,j}|^2-p|a_{s-1,p}|^2]$ with $s=1,\ldots,m$ and $p=1,\ldots,m-1$, and let
$C_2$ be an $m\times \frac{m(m-1)}{2}$ matrix with the $s$-th row and $kl$-column entry given by
$\frac{1}{\sqrt{2}}(a^*_{s-1,k}a_{s-1,l}+a^*_{s-1,l}a_{s-1,k})$ with $s=1,\ldots,m$, $k,l=1,\ldots,m-1$, and $k \neq l$. Finally,
let $C_3$ be an $m\times \frac{m(m-1)}{2}$ matrix with the $s$-th row and $kl$-th column entry given by
$\frac{i}{\sqrt{2}}(a^*_{s-1,k}a_{s-1,l}-a^*_{s-1,l}a_{s-1,k})$, with $s=1,\ldots,m$, $k,l=1,\ldots,m-1$, and $k \neq l$.
Then, the $m\times (m^2-1)$ matrix $C$ defined by the block matrices $C_1$, $C_2$, and $C_3$ as $C=(C_1, C_2, C_3)$ has rank $m-1$.

The proof of the lemma is given in the Appendix. From this lemma, we can prove the following result:

{\bf Theorem 1}\label{th1-rank of matrix}
Suppose that ${\cal M}$ is an arbitrary von Neumann measurement acting on an $m$-dimensional quantum system. Then, the $(m^2-1)\times (m^2-1)$
matrix $M$ defined in Eq. (\ref{operation and matrix}) is of rank $m-1$.

[{\sf Proof}].~ The generalized Gell Mann basis for an $m$-dimensional quantum system can be constructed as follows. For $1\leq p \leq m-1$,
\begin{eqnarray}\label{a generator-1}
w_p=\sqrt{\frac{1}{p(p+1)}} (\sum_{a=0}^{p-1} |a\rangle \langle a| - p |p\rangle \langle p|),
\end{eqnarray}
and
\begin{eqnarray}\label{a generator-2}\nonumber
w_{kl}&=&\frac{1}{\sqrt{2}}(|k\rangle \langle l|+|l\rangle \langle k|),\\
w_{kl}^\prime&=&\frac{i}{\sqrt{2}}(|k\rangle \langle l|-|l\rangle \langle k|),
\end{eqnarray}
with $k<l$, $k,l=0,1,\ldots,m-1$. In the following, we write
$w_p=w_{kl}$ for $p=m,\ldots,m+\frac{m(m-1)}{2}$ and $w_p=w^\prime_{kl}$ for $p=m+1+\frac{m(m-1)}{2},\ldots,m^2-1$, for simplicity.
Therefore, $\{w_p\}_{p=1}^{m-1}\cup\{w_{kl},w_{kl}^\prime \}_{k,l=0,k<l}^{m-1}=\{w_p\}_{p=1}^{m^2-1}$ is an orthonormal basis of the $m\times m$ traceless Hermitian operators.

First, we show that for the orthonormal basis of $m\times m$ traceless Hermitian matrices given in Eq. (\ref{a generator-1}) and Eq. (\ref{a generator-2}),
the matrix $M$ corresponding to a von Neumann measurement ${\cal M}$ has rank $m-1$. In fact, by performing a straightforward calculation we have that
\begin{widetext}
\begin{eqnarray*}
{\cal M}(w_p)&=&\sqrt{\frac{1}{p(p+1)}}[(\sum_{j=0}^{p-1} |a_{0j}|^2-p|a_{0,p}|^2)|\phi_0\rangle\langle\phi_0| + (\sum_{j=0}^{p-1} |a_{1j}|^2-p|a_{1,p}|^2)|\phi_1\rangle\langle\phi_1|\\&&
+\ldots +(\sum_{j=0}^{p-1} |a_{m-1,j}|^2-p|a_{m-1,p}|^2)|\phi_{m-1}\rangle\langle\phi_{m-1}|],\\
{\cal M}(w_{kl})&=&\frac{1}{\sqrt{2}}[(a^*_{0k}a_{0l}+a^*_{0l}a_{0k})|\phi_0\rangle\langle\phi_0|+ (a^*_{1k}a_{1l}+a^*_{1l}a_{1k})|\phi_1\rangle\langle\phi_1|\\&&
+\ldots +(a^*_{m-1,k}a_{m-1,l}+a^*_{m-1,l}a_{m-1,k})|\phi_{m-1}\rangle\langle\phi_{m-1}|],\\
{\cal M}(w^\prime_{kl})&=&\frac{i}{\sqrt{2}}[(a^*_{0k}a_{0l}-a^*_{0l}a_{0k})|\phi_0\rangle\langle\phi_0|+ (a^*_{1k}a_{1l}-a^*_{1l}a_{1k})(|\phi_1\rangle\langle\phi_1|\\&&
+\ldots +(a^*_{m-1,k}a_{m-1,l}-a^*_{m-1,l}a_{m-1,k})|\phi_{m-1}\rangle\langle\phi_{m-1}|],
\end{eqnarray*}
where $p=1,\ldots,m-1$, $k,l=0,\ldots,m-1$, and $k\neq l$.
Therefore,
\begin{eqnarray}
&&{\cal M}(w_1,\ldots,w_{m-1},w_m,\ldots,w_{m+\frac{m(m-1)}{2}},
w_{m+1+\frac{m(m-1)}{2}},\ldots,w_{m^2-1})\nonumber\\[4mm]
&&=(|\phi_0\rangle\langle\phi_0|,
|\phi_1\rangle\langle\phi_1|,\ldots,
|\phi_{m-1}\rangle\langle\phi_{m-1}|)C  \label{measurement C}\\[4mm]
&&=(w_1,\ldots,w_{m-1},w_m,\ldots,w_{m+\frac{m(m-1)}{2}},
w_{m+1+\frac{m(m-1)}{2}},\ldots,w_{m^2-1})M^\prime,\label{measurement M}
\end{eqnarray}
\end{widetext}
where the $m\times (m^2-1)$ block matrix $C=(C_1, C_2, C_3)$ is defined as in Lemma.

By applying the Lemma, the rank of the matrix $C$ is $m-1$. Hence, the maximum  of linearly independent combinations of $m^2-1$ elements from $\{|\phi_0\rangle\langle\phi_0|,
|\phi_1\rangle\langle\phi_1|,\ldots,
|\phi_{m-1}\rangle\langle\phi_{m-1}|\}$
in Eq. (\ref{measurement C}) is $m-1$. This implies that the maximum  of corresponding linearly independent combinations of $m^2-1$ elements of $\{w_1,\ldots,w_{m-1},w_m,\ldots,w_{m+\frac{m(m-1)}{2}},
w_{m+1+\frac{m(m-1)}{2}},\ldots,w_{m^2-1}\}$
in Eq. (\ref{measurement M}) is $m-1$. Therefore, the rank of $M^\prime$ is also $m-1$, i.e.,
the rank of the matrix $M^\prime$ corresponding to ${\cal M}$ under the basis given by Eq. (\ref{a generator-1}) and Eq. (\ref{a generator-2}) is $m-1$.

An arbitrary orthonormal basis of the $m\times m$ traceless Hermitian matrices $\{\mu_i\}_{i=1}^{m^2-1}$ can be obtained from
the basis given by Eq. (\ref{a generator-1}) and Eq. (\ref{a generator-2}) as $(\mu_1,\ldots,\mu_{m^2-1})=(w_1,\ldots,w_{m^2-1})O$, where
$O$ is an orthogonal matrix. If $M$ is the matrix corresponding to ${\cal M}$ under the basis $\{\mu_i\}_{i=1}^{m^2-1}$,
then the matrix $M^\prime$ corresponding to ${\cal M}$ under the basis $\{w_i\}_{i=1}^{m^2-1}$ is $OMO^{T}$.
Because the rank of the matrix $M$ corresponding to ${\cal M}$ is invariant under the transformation of the orthogonal matrix,
the rank of the matrix $M$ in Eq. (\ref{operation and matrix}) is $m-1$.
\qed

In a two dimensional quantum system, suppose that a von Neumann measurement ${\cal M}$ is the projector on the computational basis, ${\cal M}=\{|0\rangle\langle0|,|1\rangle\langle1|\}$. If we choose the orthonormal basis of $2\times2$ traceless Hermitian matrices given by the Pauli matrices
\begin{eqnarray*}
\sigma_1=\left(
\begin{array}{cc}
0& 1\\
1 & 0
\end{array}
\right),\ \ \sigma_2=\left(
\begin{array}{cc}
0& -{\rm i}\\
{\rm i} & 0
\end{array}
\right),\ \ \sigma_3=\left(
\begin{array}{cc}
1& 0\\
0 & -1
\end{array}
\right),
\end{eqnarray*}
then the matrix corresponding to the von Neumann measurement ${\cal M}$ is $M=\left(
\begin{array}{ccc}
0& 0&0\\
0& 0&0\\
0 & 0 &1
\end{array}
\right)$, which is of rank one.

In a three dimensional quantum system, suppose that a von Neumann measurement ${\cal M}$ is the projector on the computational basis, ${\cal M}=\{|0\rangle\langle0|,|1\rangle\langle1|,|2\rangle\langle2|\}$. If we choose the orthonormal basis of $3\times3$ traceless Hermitian matrices as the Gell-mann matrices
\begin{eqnarray*}
\lambda_1=\left(
\begin{array}{ccc}
0& 1&0\\
1 & 0&0\\
0 & 0&0\\
\end{array}
\right),\ \ \lambda_2=\left(
\begin{array}{ccc}
0& -{\rm i}&0\\
{\rm i} & 0&0\\
0 & 0&0\\
\end{array}
\right),\ \ \lambda_3=\left(
\begin{array}{ccc}
1& 0&0\\
0 & -1&0\\
0 & 0&0\\
\end{array}
\right),\ \ \lambda_4=\left(
\begin{array}{ccc}
0& 0&1\\
0 & 0&0\\
1 & 0&0\\
\end{array}
\right),
\\ \lambda_5=\left(
\begin{array}{ccc}
0& 0&-{\rm i}\\
0 & 0&0\\
{\rm i} & 0&0\\
\end{array}
\right),\ \ \lambda_6=\left(
\begin{array}{ccc}
0& 0&0\\
0 & 0&1\\
0 & 1&0\\
\end{array}
\right),\ \ \lambda_7=\left(
\begin{array}{ccc}
0& 0&0\\
0 & 0&-{\rm i}\\
0 & {\rm i}&0\\
\end{array}
\right),\ \ \lambda_8=\frac{1}{\sqrt{3}}\left(
\begin{array}{ccc}
1& 0&0\\
0 & 1&0\\
0 & 0&-2\\
\end{array}
\right),
\end{eqnarray*}
then the matrix corresponding to the von Neumann measurement ${\cal M}$ is the $8\times8$ diagonal matrix ${\rm diag}(0,0,1,0,0,0,0,1)$, which is of rank two.

We remark that because the basis $\{\mu_i\}_{i=1}^{m^2-1}$ of $m\times m$ traceless Hermitian matrices is not necessarily orthonormal,
the matrix $M$ corresponding to the von Neumann measurement ${\cal M}$ given in Eq. (\ref{operation and matrix}) is still of rank $m-1$.
This can be seen from the proof of Theorem 1, because replacing the orthogonal matrix by a reversible matrix does not change the rank.

Moreover, for any $(m^2-1)\times (m^2-1)$ idempotent matrix $Y$ of rank $m-1$, there must exist a basis of $m\times m$
traceless Hermitian matrices such that $Y$ corresponds to the von Neumann measurement ${\cal M}$.
In fact, for any idempotent matrix $Y$ of rank $m-1$, there exists a reversible matrix $Q$ such that $Y=QMQ^{-1}$.
Hence, if $M$ is the matrix corresponding to the von Neumann measurement ${\cal M}$ under the basis $\{\mu_i\}_{i=1}^{m^2-1}$,
then $Y$ is the matrix corresponding to the von Neumann measurement ${\cal M}$ under the basis $\{\mu^\prime_i\}_{i=1}^{m^2-1}$
defined by $(\mu^\prime_1,\ldots,\mu^\prime_{m^2-1})=(\mu_1,\ldots,\mu_{m^2-1})Q^{-1}$.

\section{Necessary conditions for nullity of quantum correlation}

By utilizing Theorem 1, we obtain necessary conditions for the nullity of quantum correlation as follows.

{\bf Theorem 2}
Consider an $m\otimes n$ $(m\leq n)$ quantum state in the Bloch representation:
\begin{widetext}
\begin{eqnarray}\label{bipartite bloch}
\rho=\frac{1}{mn}(I_m\otimes I_n +\sum_{i=1}^{m^2-1} r_i \mu_i\otimes I_n
+\sum_{j=1}^{n^2-1} s_j I_m\otimes \nu_j +\sum_{i=1}^{m^2-1}\sum_{j=1}^{n^2-1} t_{ij}\mu_i\otimes \nu_j).
\end{eqnarray}
\end{widetext}
If it is classical-quantum correlated, then the rank of $(R,T)$ is not greater than $m-1$.
If it is quantum-classical correlated, then the rank of $(S, T^T)$ is not greater than $n-1$. If it is classical-classical correlated, then the rank of $\left(
\begin{array}{cc}
1&S^T\\
R&T
\end{array}
\right)$ is not greater than $m$.
Here, $R=(r_1, r_2,\ldots,r_{m^2-1})^T$ and $S=(s_1, s_2,\ldots,s_{n^2-1})^T$ are the vectors with components $r_i$ and $s_j$, $T=(t_{ij})$ is the matrix with entries $t_{ij}$, $I_m$ is the $m\times m$ identity matrix, $I_n$ is the $n\times n$ identity matrix, and
$\{\mu_i\}_{i=1}^{m^2-1}$ and $\{\nu_j\}_{j=1}^{m^2-1}$ are the orthonormal bases of the $m\times m$ and $n\times n$ traceless Hermitian matrices, respectively.

[{\sf Proof}].~If $\rho$ is classical-quantum correlated, then
there exists a von Neumann measurement ${\cal M}$ acting on the first subsystem
such that $({\cal M}\otimes I)\rho({\cal M}^\dagger\otimes I)=\rho$. Note that
\begin{widetext}
\begin{eqnarray}
&&({\cal M}\otimes I)\rho({\cal M}^\dagger\otimes I)\\\nonumber
&&=\displaystyle
\frac{1}{mn}[I\otimes I+\sum_{i=1}^{m^2-1} (M R)_i \,\mu_i\otimes I+ \sum_{j=1}^{n^2-1} s_j I\otimes \nu_j +\sum_{i=1}^{m^2-1}\sum_{j=1}^{n^2-1} (M T)_{ij}\,\mu_i\otimes\nu_j],
\end{eqnarray}
\end{widetext}
by Eq. (\ref{operation and matrix}), where $({\cal M}\otimes I) \rho ({\cal M}^\dagger\otimes I)=\rho$ indicates that $M R=R$ and $M T=T$.
Because $M$ has at most $m-1$ independent eigenvectors by Theorem 1, it follows that the rank of the matrix $(R,T)$ is not greater than $m-1$. Similarly, if $\rho$ is quantum-classical correlated, then there exists a von Neumann measurement ${\cal M^\prime}$ acting on the second subsystem such that $(I \otimes {\cal M}^\prime) \rho (I \otimes {\cal M}^{\prime\dagger})=\rho$.
Note that
\begin{widetext}
\begin{equation}
\begin{array}{rcl}
&&(I \otimes {\cal M}^\prime) \rho (I \otimes {\cal M}^{\prime\dagger})\\
&&=\displaystyle
\frac{1}{mn}[I\otimes I+\sum_{i=1}^{m^2-1} r_i \,\mu_i\otimes I+ \sum_{j=1}^{n^2-1} (M^{\prime} S)_j I\otimes \nu_j +\sum_{i=1}^{m^2-1}\sum_{j=1}^{n^2-1} (T(M^{\prime})^T)_{ij}\,\mu_i\otimes\nu_j],
\end{array}
\end{equation}
\end{widetext}
by Eq. (\ref{operation and matrix}), where $(I \otimes {\cal M}^\prime) \rho (I \otimes {\cal M}^{\prime\dagger})=\rho$ indicates that $M^\prime S=S$ and $M^\prime T^T=T^T$.
Because $M^\prime$ has at most $n-1$ independent eigenvectors, by Theorem 1, it follows that the rank of the matrix $(S, T^T)$, or $\left(
\begin{array}{cc}
S^T\\
T
\end{array}
\right)$, is not greater than $n-1$.
Finally, if
$\rho$ is classical-classical correlated,
then it satisfies the equalities $M R=R$, $M T=T$, $M^\prime S=S$, and $M^\prime T^T=T^T$. Thus, we obtain that the rank of $\left(
\begin{array}{cc}
1&S^T\\
R&T
\end{array}
\right)$ is not greater than $m$.
\qed

Theorem 2 provides necessary conditions for the nullity of quantum correlation in terms of the coefficients in the Bloch representation. Here, the matrix $\left(
\begin{array}{cc}
1&S^T\\
R&T
\end{array}
\right)$ is in fact the correlation matrix in Ref. \cite{B. Dakic}. In Ref. \cite{B. Dakic}, it is proved that if a quantum state is classical-quantum correlated, then the rank of $\left( \begin{array}{cc}
1&S^T\\
R&T
\end{array}
\right)$ is not greater than $m$. Here, Theorem 2 shows that if a quantum state is classical-quantum correlated, then the rank $(R,T)$ is not greater than $m-1$, which is sufficient for the rank of $\left( \begin{array}{cc}
1&S^T\\
R&T
\end{array}
\right)$ to be no greater than $m$. Therefore, as a necessary condition for the nullity of a quantum correlation, Theorem 2 is stronger than the condition in Ref. \cite{B. Dakic}.
For example, let us consider a quantum state
\begin{eqnarray}
\rho_0=\frac{1}{4}[I_2\otimes I_2 +\frac{1}{x} \sigma_3\otimes I_2 + \frac{1}{x}I_2\otimes \sigma_3+\frac{1}{x^2}\sigma_1\otimes\sigma_1+\frac{1}{x^2} \sigma_3\otimes\sigma_3]
\end{eqnarray}
with $x\geq \sqrt{2}$. For this quantum state, one can obtain that $R=S=(0,0,\frac{1}{x})^T$, with $T={\rm diag}(\frac{1}{x^2},0,\frac{1}{x^2})$. Then, $(R,T)=\left( \begin{array}{cccc}
0& \frac{1}{x^2} & 0&0\\
0& 0&0&0\\
\frac{1}{x} & 0&0&\frac{1}{x^2}
\end{array}
\right)$ is of rank two, which implies by Theorem 2 that $\rho_0$ is not classical-quantum correlated. Similarly, we obtain that $(S, T^T)$ is of rank two, and so $\rho_0$ is not quantum-classical. However, the correlation matrix $\left( \begin{array}{cc}
1&S^T\\
R&T
\end{array}
\right)=\left( \begin{array}{cccc}
1&0&0&\frac{1}{x}\\
0& \frac{1}{x^2} & 0&0\\
0& 0&0&0\\
\frac{1}{x} & 0&0&\frac{1}{x^2}
\end{array}
\right)$ is also of rank two. By the nullity condition in Ref. \cite{B. Dakic}, we cannot determine the quantum correlation in the quantum state $\rho_0$.

Now as another example, we consider two-qubit states with maximally mixed marginals, which are referred as Bell diagonal states. Such states are locally equivalent to
$$
\rho_1=\frac{1}{4}(I+\sum_{i=1}^3 t_i \sigma_i \otimes \sigma_i),
$$
where $\sigma_i$ are the Pauli matrices, and
$\vec{t}=(t_1,t_2,t_3)$ belongs to the tetrahedron defined by the set of vertices $(-1,-1,-1)$, $(-1,1,1)$, $(1,-1,1)$, and $(1,1,-1)$.
Here, $\rho_1$ is separable if $\vec{t}$ belongs to the octahedron defined by the set of vertices $(\pm1,0,0)$, $(0,\pm1,0)$, and $(0,0,\pm1)$.
For a Bell diagonal state that is already expressed in the Bloch representation, it is easy to obtain $R=0$, $S=0$, and $T={\rm diag}(t_1,t_2,t_3)$. If there is more than one nonzero element in $\{t_1,t_2,t_3\}$, then the ranks of $(R,T)$ and $(S, T^T)$ are greater than one. By Theorem 2, this class of Bell diagonal states are not classical-quantum or quantum-classical. This means they are quantum-quantum correlated. Meanwhile, the only Bell diagonal states that have no more than one nonzero element in $\{t_1,t_2,t_3\}$ are the vertices and the central point. These seven quantum states are indeed classical-classical correlated \cite{S. Weis}.
Therefore, a necessary and sufficient condition for $\rho_1$ to be a quantum-quantum state is that
the rank of $T={\rm diag}(t_1,t_2,t_3)$ is greater than one.
More generally, for an $m\otimes m$ quantum state
$$
\rho_2=\frac{1}{m^2}(I_m\otimes I_m+\sum_{i=1}^{m^2-1}t_i\mu_i\otimes\mu_i),
$$
with $\{\mu_i\}$ representing the orthonormal basis of $m\times m$ traceless Hermitian matrices,
it follows from Theorem 2 that if there are more than $m-1$ nonzero elements in $\{t_i\}$, then $\rho_2$ is a quantum-quantum correlated state.

\begin{center}
\begin{figure}
\label{fig}
\resizebox{6cm}{!}{\includegraphics{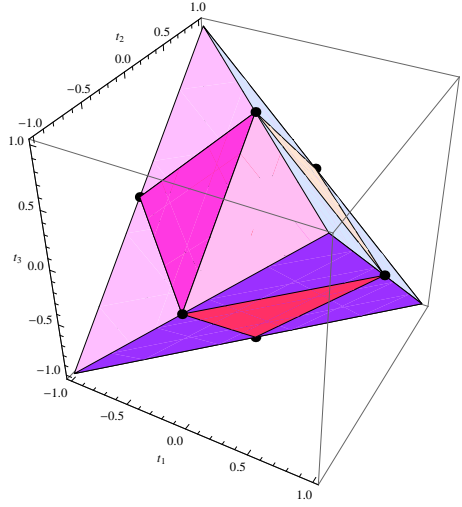}}
\caption{(Color online) Geometry of Bell-diagonal states. }
\end{figure}
\end{center}

\section{Conclusions}

We have investigated the properties of matrices related to the von Neumann measurement, and the nullity condition for quantum correlation.
It has been demonstrated that these matrices are idempotent, and have rank $m-1$.
Based on this result, we have obtained three necessary conditions for the nullity of quantum correlations in bipartite systems.
Our results may shed light on the properties of von Neumann measurements, and can be used to detect quantum correlation.

\section{Appendix}

{\sf Proof of Lemma}~ By performing some elementary matrix transformations, we find that the rank of the matrix $C$ is equal to the rank of the matrix $C_0$,
\begin{eqnarray*}
C_0=\left(
\begin{array}{cccccccc}
\cdots & |a_{00}|^2-|a_{0i}|^2 & \cdots & a^*_{0k}a_{0l} & \cdots  \\
 \cdots & |a_{10}|^2-|a_{1i}|^2 & \cdots & a^*_{1k}a_{1l} & \cdots  \\
\cdots & \cdots & \cdots & \cdots & \cdots \\
 \cdots & |a_{m-1,0}|^2-|a_{m-1,i}|^2 & \cdots & a^*_{m-1,k}a_{m-1,l} & \cdots
\end{array}
\right),
\end{eqnarray*}
which is composed of the column vectors
$$\vec{\alpha}_i=(|a_{00}|^2-|a_{0i}|^2, |a_{10}|^2-|a_{1i}|^2, \ldots, |a_{m-1,0}|^2-|a_{m-1,i}|^2)^T $$
for $i=1,\ldots,m-1$, and
$$\vec{\beta}_{kl}=(a^*_{0k}a_{0l},a^*_{1k}a_{1l}, \ldots, a^*_{m-1,k}a_{m-1,l})^T$$
for $k,l=0,\ldots,m-1$ and $k\neq l$.

To show that the rank of $C_0$ is $m-1$, let us first consider the case that $m=2$,
\begin{eqnarray*}
C_0=\left(
\begin{array}{cccccccc}
|a_{00}|^2-|a_{01}|^2  & a^*_{00}a_{01} & a^*_{01}a_{00} \\
|a_{10}|^2-|a_{11}|^2  & a^*_{10}a_{11} & a^*_{11}a_{10}  \\
\end{array}
\right).
\end{eqnarray*}
Here, $C_0$ is of rank one, because the sum of the two rows is zero, as $A^\dagger A=I_2$. Hence, the stated conclusion is true for $m=2$. Now, suppose that the conclusion is also true for all $k\times k$ unitary matrices $A$ with $k<m$. Noting that the sum of the $m$ rows of the matrix $C_0$ is zero, it follows that the rank of $C_0$ is no greater than $m-1$. Next, we show that there are $m-1$ linearly independent column vectors in $C_0$.

Note that the order of the column vectors in $A$ has no effect on the rank of the matrix $C$. Without loss of generality, we suppose that the first $p_0$ elements of the first column of $A$
are all nonzero, and the rest are all zero. Then, $1\leq p_0\leq m$. If $p_0=1$, then $A$ is a direct sum of $1$ and an $(m-1)\times (m-1)$ unitary matrix $A^\prime$.
By assumption, because the matrix $C_0^\prime$ is deduced from $A^\prime$ in the same manner as $C_0$ is deduced
from $A$, it is rank of $m-2$, and there exist $m-2$ linearly independent column vectors $\vec{\gamma}_1^\prime,\vec{\gamma}_2^\prime,\ldots,\vec{\gamma}_{m-2}^\prime$ in $C_0^\prime$.
Therefore, the $m-1$ linearly independent vectors of $C_0$ are $(0,\vec{\gamma}_1^{\prime T})^T,(0,\vec{\gamma}_2^{\prime T})^T,\ldots,(0,\vec{\gamma}_{m-2}^{\prime T})^T$ and the first column vector $\vec{\alpha}_1=(1,-|a_{11}|^2,\ldots,-|a_{m-1,1}|^2)^T$ in $C_0$.

For $p_0>1$, consider the column vectors $\vec{\beta}_{01}, \vec{\beta}_{02}, \ldots, \vec{\beta}_{0,{m-1}}$.
Let $x_1\vec{\beta}_{01}+x_2\vec{\beta}_{02}+\ldots+x_{m-1}\vec{\beta}_{0,{m-1}}=0$. Then, one has a set of linear equations
\begin{equation}\label{system eq in step 1}
\begin{array}{rcl}
x_1a_{01}+x_2a_{02}+\ldots +x_{m-1}a_{0,{m-1}}&=&0,\\
x_1a_{11}+x_2a_{12}+\ldots +x_{m-1}a_{1,{m-1}}&=&0,\\
&\vdots&\\
x_1a_{p_0-1,1}+x_2a_{p_0-1,2}+\ldots +x_{m-1}a_{p_0-1,{m-1}}&=&0.
\end{array}
\end{equation}
Taking into account that the rank of the $p_0\times m$ matrix
\begin{eqnarray*}
F_1=\left(
\begin{array}{ccccc}
a_{00} & a_{01}& a_{02} & \cdots & a_{0,{m-1}} \\
a_{10} & a_{11}& a_{12} & \cdots & a_{1,{m-1}} \\
\cdots & \cdots & \cdots &\cdots & \cdots \\
a_{p_0-1,0} & a_{p_0-1,1}& a_{p_0-1,2} & \cdots & a_{p_0-1,{m-1}}
\end{array}
\right)
\end{eqnarray*}
is $p_0$, because of the unitarity of $A$, we have that the maximum  of linearly independent column vectors of $F_1$ is $p_0$.
Because the first column vector of $A$ is linearly independent from the others, the rank of the coefficient matrix
in Eqs. (\ref{system eq in step 1}) is $p_0-1$. This demonstrates that the maximum  of linearly independent
column vectors $\{\vec{\beta}_{01}, \vec{\beta}_{02} \ldots, \vec{\beta}_{0,{m-1}}\}$ is $p_0-1$. Without loss of generality, suppose that
\begin{equation}
\begin{array}{rcl}
\vec{\gamma}_1 &:=&
(a^*_{00}a_{0,i_1},\ldots,a^*_{p_0-1,0}a_{p_0-1,i_1},0,\ldots,0)^T,\\
\vec{\gamma}_2 &:=&
(a^*_{00}a_{0,i_2},\ldots,a^*_{p_0-1,0}a_{p_0-1,i_2},0,\ldots,0)^T, \\
&\vdots& \\
\vec{\gamma}_{p_0-1} &:=&
(a^*_{00}a_{0,i_{p_0-1}},\ldots,a^*_{p_0-1,0}a_{p_0-1,i_{p_0-1}},0,\ldots,0)^T
\end{array}
\end{equation}
are these linearly independent vectors.

Now, we consider the column vectors of $A$ other than the first,
and pick out the column vector with the most nonzero entries between the $p_0+1$-th entry and the $m$-th entry.
Without loss of generality, we suppose that this is the second column vector of $A$, and it has $p_1$ nonzero entries between the $p_0+1$-th and $p_0+p_1$-th positions,
with $1\leq p_1\leq m-p_0$. If $p_1=1$, then
$\vec{\alpha}_1=(|a_{00}|^2-|a_{01}|^2, \ldots, |a_{p_0-1,0}|^2-|a_{p_0-1,1}|^2, -|a_{p_0,1}|^2, 0,\ldots,0 )^T$
is linearly independent from the vectors $\vec{\gamma}_{1}$, $\vec{\gamma}_{2}$, $\ldots$, $\vec{\gamma}_{p_0-1}$.

If $p_1>1$, then we check the column vectors $\vec{\beta}_{10}$, $\vec{\beta}_{12}$, $\ldots$, $\vec{\beta}_{1,{m-1}}$ and
$\vec{\alpha}_1$.
Let $x_0\vec{\beta}_{10}+x_1\vec{\alpha}_1+x_2\vec{\beta}_{12}+\ldots+x_{m-1}\vec{\beta}_{1,{m-1}}=0$. Then, we have the linear equations
\begin{widetext}
\begin{equation}\label{system eq in step 2}
\begin{array}{rcl}
x_0a_{p_0,0}-x_1a_{p_0,1}+x_2a_{p_0,2}+x_3a_{p_0,3}+\ldots +x_{m-1}a_{p_0,{m-1}}&=&0,\\
x_0a_{p_0+1,0}-x_1a_{p_0+1,1}+x_2a_{p_0+1,2}+x_3a_{p_0+1,3}+\ldots +x_{m-1}a_{p_0+1,{m-1}}&=&0,\\
&\vdots&\\
x_0a_{p_0+p_1-1,0}-x_1a_{p_0+p_1-1,1}+x_2a_{p_0+p_1-1,2}+x_3a_{p_0+p_1-1,3}+\ldots +x_{m-1}a_{p_0+p_1-1,{m-1}}&=&0.
\end{array}
\end{equation}
Taking into account that the rank of the coefficient matrix
\begin{eqnarray*}
F_2=\left(
\begin{array}{ccccc}
a_{p_0,0} & -a_{p_0,1} & a_{p_0,2} & \cdots & a_{p_0,{m-1}}\\
a_{p_0+1,0} & -a_{p_0+1,1} & a_{p_0+1,2} &\cdots & a_{p_0+1,{m-1}}\\
\cdots & \cdots & \cdots &\cdots &\cdots \\
a_{p_0+p_1-1,0} & -a_{p_0+p_1-1,1} & a_{p_0+p_1-1,2} & \cdots & a_{p_0+p_1-1,{m-1}}
\end{array}
\right)
\end{eqnarray*}
\end{widetext}
from Eqs. (\ref{system eq in step 2}) is $p_1$, because of the unitarity of $A$, there exist $p_1$ linearly independent vectors in $\vec{\beta}_{10}, \vec{\beta}_{12}, \ldots, \vec{\beta}_{1,{m-1}}$ and $\vec{\alpha}_1$.
Without loss of generality, we assume that
\begin{equation}
\begin{array}{rcl}
\vec{\gamma}_{p_0}&=&(y_{1,p_0},\ldots,y_{p_0+p_1-1,p_0},0,\ldots,0)^T,
\\ \vec{\gamma}_{p_0+1}&=&(y_{1,p_1},\ldots,y_{p_0+p_1-1,p_1},0,\ldots,0)^T, \\
&\vdots& \\
\vec{\gamma}_{p_0+p_1-1}&=&(y_{1,p_0+p_1-1},\ldots,y_{p_0+p_1-1,p_0+p_1-1},0,\ldots,0)^T
\end{array}
\end{equation}
are linearly independent with $(y_{p_0,j},\ldots,y_{p_0+p_1-1,j})\neq 0$ for $j=p_0, \ldots, p_0+p_1-1$. It is obvious that these $p_1$ vectors $\vec{\gamma}_{p_0},\ldots,\vec{\gamma}_{p_0+p_1-1}$
are linearly independent from the vectors $\vec{\gamma}_{1},\ldots,\vec{\gamma}_{p_0-1}$. Therefore, we now have $p_0+p_1-1$ linearly independent column vectors in the matrix $C_0$.

Finally, we check the column vectors of $A$ other than the first two,
and pick out the column vector that has the most nonzero entries between the $(p_0+p_1)$-th entry and the $m$-th entry.
By continuing the procedure above, we can find $m-1$ linearly independent vectors.
\qed

\section{Acknowledgement}

This work is supported by the NSF of China
under Grant No. 11401032 and the NSF of Hainan Province under Grant No. 20161006.

\end{document}